\documentclass[aps,prd,floats,onecolumn,nofootinbib]{revtex4}
\pdfoutput=1  
\usepackage{graphicx,color}
\usepackage{subfigure}
\usepackage{latexsym,amsmath,amssymb,graphicx,booktabs}
\usepackage{hyperref}

\definecolor{MyBlue}{rgb}{0.15,0.15,0.70}

\hypersetup{
colorlinks=true,
citecolor=MyBlue,
linkcolor=MyBlue,
urlcolor=MyBlue
}

\usepackage{amssymb}
\usepackage{amsmath}
\usepackage{amsfonts}
\usepackage{upgreek}
\usepackage{latexsym}

 \usepackage{slashed}
\usepackage{slantsc}
\usepackage{braket}
\usepackage{mathrsfs}
\usepackage{placeins}

\newcommand{\bk}{{\mathbf k}}

\newcommand{\bq}{{\mathbf q}}

\newcommand{\A}{{\mathcal A}}

\newcommand{\HH}{{\cal H}}

\newcommand{\De}{\Delta}

\newcommand{\la}{\lambda}
\newcommand{\si}{\sigma}

\newcommand{\bea}{\begin{eqnarray}}
\newcommand{\eea}{\end{eqnarray}}
\newcommand{\bean}{\begin{eqnarray*}}
\newcommand{\eean}{\end{eqnarray*}}

\newcommand{\ssu}{\slashed \partial}
\newcommand{\ssd}{\slashed \partial^*}

\newcommand{\td}{\text{d}}

\def\id{{\rm 1\kern -2.5pt I}}

\graphicspath{ {./Graphs/} }

\include{mydefs}

\begin{document}

\vspace*{2cm}

\title{Lensing signals from Spin-2 perturbations}
\author{Julian Adamek}
\email{julian.adamek@unige.ch}
\affiliation{D\'epartement de Physique Th\'eorique \& Center for Astroparticle Physics, Universit\'e de Gen\`eve, 24 quai E.~Ansermet, CH--1211 Gen\`eve 4, Switzerland}
\author{Ruth Durrer}
\email{ruth.durrer@unige.ch}
\affiliation{D\'epartement de Physique Th\'eorique \& Center for Astroparticle Physics, Universit\'e de Gen\`eve, 24 quai E.~Ansermet, CH--1211 Gen\`eve 4, Switzerland}
\author{Vittorio Tansella}
\email{vittorio.tansella@unige.ch}
\affiliation{D\'epartement de Physique Th\'eorique \& Center for Astroparticle Physics, Universit\'e de Gen\`eve, 24 quai E.~Ansermet, CH--1211 Gen\`eve 4, Switzerland}

\date{\today}

\begin{abstract}
We compute the angular power spectra of the E-type and B-type lensing potentials for gravitational waves from inflation and for 
 tensor perturbations induced by scalar perturbations. We derive the tensor-lensed CMB power spectra for both cases. We also apply our formalism to determine the linear lensing potential for a Bianchi~I spacetime with small anisotropy.
\end{abstract}

\maketitle
\flushbottom

\section{Introduction}
\label{s:intro}
Since many years now, the cosmic microwave background (CMB) is the most precious signal in cosmology. It is being used to determine the parameters which govern the expansion of the Universe, its content and the initial conditions of its fluctuations, see~\cite{Ade:2015xua} for latest results, \cite{Durrer:2015aa} for a historical account and~\cite{Ruth_book} for a comprehensive monograph on the subject. So far, no gravitational wave background has been detected in the CMB and an upper limit of $r\lesssim 0.1$ has been derived for the tensor to scalar ratio~\cite{Ade:2015tva}. In cosmology, these parameter estimations are of course always model dependent and therefore have to be taken with a grain of salt. The present limit on $r$ mainly comes from the contribution of gravitational waves to the temperature anisotropy and from that fact that they induce so called $B$-polarisation, i.e., a rotational component in the polarisation vector field which is absent for purely scalar perturbations. In this paper we study yet another aspect of tensor perturbations. They introduce B-modes (i.e.\ rotational modes)  also in the lensing signal. 
This effect has already been derived in~\cite{Dodelson:2003bv,Li:2006si,Yamauchi:2012bc} and applied to both, primordial gravitational waves from inflation and topological defects~\cite{Yamauchi:2012bc,Yamauchi:2013fra}. In both cases it was found that for realistic parameters the effect is unobservably small. 
Also the effect of the gravitational wave contributions to the shear in the lensing of large scale structure has been investigated and found to be very small~\cite{Schmidt:2012ne,Schmidt:2012nw}, while the effect of the tidal field seems to be more promising~\cite{Schmidt:2013gwa}. In this work we present an independent, alternative derivation of CMB lensing by tensor modes. As we will see, these modes can describe classical gravitational waves but also other spin-2 perturbations of the metric which do not propagate as waves in the usual sense, e.g.\ extremely infrared modes which make the Universe locally look like a Bianchi I model. We compare the signal induced by primordial gravitational waves and the one from tensor perturbations which are generated by the formation of large scale structure. This part is new. In a previous paper, the effect from induced tensor perturbations has been considered to second order~\cite{Saga:2015apa}, while here we incorporate the fully nonperturbative results from relativistic large scale structure simulations~\cite{Adamek:2014xba,Adamek:2015eda}. We also discuss the qualitative difference of these tensor modes from ordinary gravitational waves which has not been appreciated in previous literature. Finally, we apply our formalism also to describe lensing in a Bianchi I model, i.e.\ a homogeneous but anisotropic model of the Universe.

The paper is structured as follows: In Section~\ref{s:der} we derive the main formulae for CMB lensing by gravitational waves. In Section~\ref{s:num} we present numerical results for primordial gravitational waves and for gravitational waves induced from large scale structure. In Section~\ref{s:ani} we study the Bianchi I model and in Section~\ref{s:con} we discuss our findings and conclude. Some technical aspects are deferred to two appendices. 

\section{CMB lensing by gravitational waves}\label{s:der}

\textbf{Notation:} We work with a flat Friedmann background and in conformal time, $t$, such that the unperturbed metric is given by

\begin{equation}
\td \bar s^2= g_{\mu \nu} \td x^\mu \td x^\nu=a^2(t)(-\td t^2+\gamma_{ij} \td x^i \td x^j)  .
\label{eq:FRWmetric}
\end{equation}
Greek indices run from 0 to 3, latin indices $(i,j,\cdots)$ run over spacelike coordinates from 1 to 3 while latin indices $(a,b,\cdots)$ go from 1 to 2 and denote coordinates on the unit sphere of directions. Throughout this paper we adopt the spherical coordinate system for the metric of the three dimensional slices and we set the spatial curvature to zero,

\begin{equation}
\gamma_{ij} \td x^i \td x^j = \td \chi^2 + \chi^2(\td \theta^2+ \sin ^2 \theta \, \,\td \phi^2) \,,
\end{equation}
where $\chi$ is the comoving distance. The derivative with respect to $t$ will be denoted with a dot $\dot x = \td x/\td t$, so that $\HH= \dot a/a$ is the comoving Hubble parameter. $H=\HH/a$ is the physical Hubble parameter with present value $H_0=h \,100$km/sec/Mpc.

\subsection{Perturbed photon path and the lensing potentials}\label{ss:1.1}

We first want to compute the deflection of a light ray in a Friedmann universe with  spin-2 perturbations. The perturbation of the line element is given by

\begin{equation}
\td s^2= a^2(t)[-\td t^2+(\gamma_{ij} +h_{ij} )\td x^i \td x^j] ,
\label{eq:pertmetr}
\end{equation}
where the tensor modes $h_{ij}$ obey the transverse $h^{ij},{}_i=0$ and traceless $h^i {}_i =0$ conditions. Since we are only interested in deflection of a null-geodesic it is more convenient to consider the conformally related metric

\begin{equation}
\td \tilde s^2= -\td t^2+(\gamma_{ij} +h_{ij} )\td x^i \td x^j .
\label{eq:pertmetrconf}
\end{equation}

The observer is placed at $\mathbf{x}=0$ and we now consider a photon with unperturbed trajectory given by the null-geodesic $[x^\mu (\la)] =\la  [1, {\mathbf{n}}]$ where 
${\mathbf{n}}$ is the photon direction fixed by the angles $(\theta_0, \phi_0)$ and $\la$ is the affine parameter, such that, at the unperturbed level, $\td t=\td \la = -\td \chi$. The perturbed four-velocity is given by $[n^\mu(\la)]=[1+\delta n^0 (\la), {\mathbf{n}} + \delta {\mathbf{n}} (\la)]$. We define the displacement vector $\boldsymbol{\alpha}= [\theta-\theta_0, \sin \theta (\phi-\phi_0)]$. Solving the spatial parts of the geodesic equations at first order in $\delta n^\mu$ and in the perturbation, we find

\begin{equation}
\alpha_a= \int_{0}^{\chi_*} \td \chi \, \left[ \frac{h_{ra}(\chi,\theta_0,\phi_0)}{\chi}+ \frac{1}{2} \frac{\chi-\chi_*}{\chi \chi_*} \nabla_a h_{rr}(\chi,\theta_0,\phi_0) \right] ,
\label{eq:defang}
\end{equation}
where $\chi_*$ is the comoving distance at emission,  $\nabla=[\partial_\theta, (\sin \theta)^{-1} \partial_\phi]$ is the gradient on the unit sphere and Born approximation was used.\\

When considering lensing by density perturbations the first-order deflection angle can be written as the gradient of a single scalar lensing potential \cite{Lewis:2006fu}. For tensor perturbations this is no longer true and we have to decompose $\boldsymbol{\alpha}$  in its gradient-mode, i.e.\ E-mode, potential $\psi ({\mathbf{n}})$ and its curl-mode, i.e.\ B-mode, potential $\varpi ({\mathbf{n}})$. We then write

\begin{equation}
\alpha_a ({\mathbf{n}}) = \nabla_a \psi({\mathbf{n}}) + \varepsilon ^{b} {}_a \; \nabla_b \varpi({\mathbf{n}}) ,
\end{equation}
explicitly 

\begin{equation}
\alpha_\theta= \partial_\theta \psi + \frac{1}{\sin \theta} \partial_\phi \varpi 
, \qquad
\alpha_\phi=  \frac{1}{\sin \theta} \partial_\phi \psi -  \partial_\theta \varpi . 
\end{equation}
Let us also introduce the spin raising and lowering operators $\ssu$ and $\ssd$ acting on a helicity $s$ tensor field on the sphere as

\begin{equation}
\ssu_s = \Bigl( s \cot \theta - \partial_\theta - \frac{i}{\sin \theta} \partial_\phi \Bigr),
\qquad
\ssd_s = \Bigl( -s \cot \theta - \partial_\theta + \frac{i}{\sin \theta} \partial_\phi \Bigr) .
\label{eq:ssud}
\end{equation}
With these operators we can write two differential equations for the helicity 0 lensing potentials

\begin{equation}
\begin{cases}
\sqrt{2} \; \alpha_+  =- \ssd(\psi+i \varpi) \\
\sqrt{2} \; \alpha_-  =- \ssu(\psi-i \varpi) ,
\end{cases}
\label{eq:system}
\end{equation}
where we have introduced the helicity basis $ \mathbf{\alpha_{\pm}}= \frac{1}{\sqrt2} (\mathbf{\alpha}_\theta \mp i \mathbf{\alpha}_\phi)$.  Using $ \ssd\ssu+\ssd\ssu = 2\De$ and the fact that the commutator of the spin raising and lowering operators vanishes on functions, we obtain  
\begin{equation}
\begin{cases}
 \ssd\alpha_- + \ssu\alpha_+  =- \sqrt{2}\De\psi \\
 \ssd\alpha_- - \ssu\alpha_+ =- \sqrt{2}\De \varpi .
\end{cases}
\label{eq:system2}
\end{equation}

In the next section we will compute $\alpha_\pm$ in order to solve eq.~(\ref{eq:system2}) for the two lensing potentials.

\subsection{Angular power spectra}\label{ss:1.2}

It is convenient to work in Fourier space where we define

\begin{equation}
h_{ij}(\mathbf{x},t) = \int \frac{\td^3 \mathbf{k}}{(2 \pi)^3} \, \left[   h_{\mathbf{k}}^{\oplus}(t) e_{ij}^{\oplus}(\hat{\mathbf{k}}) +  h_{\mathbf{k}}^{\otimes}(t)  e_{ij}^{\otimes}(\hat{\mathbf{k}})   \right] \, e^{-i \mathbf{k \cdot x}} \;. \label{eq:fourexp} \end{equation}

Here we have introduced two time-independent polarization tensors $ e_{ij}^{\oplus}(\hat{\mathbf{k}})$ and $  e_{ij}^{\otimes}(\hat{\mathbf{k}})$ which can be expressed in terms of vectors of the orthonormal basis $( \hat{\mathbf{k}},\mathbf{e}_1, \mathbf{ e}_2)$ as

\begin{equation}
e_{ij}^{\oplus}(\hat{\mathbf{k}})= \frac{1}{\sqrt{2}} [ e_i^1 (\hat{\mathbf{k}}) e_j^1 (\hat{\mathbf{k}})-   e_i^2 (\hat{\mathbf{k}})  e_j^2 (\hat{\mathbf{k}})] , \quad
 e_{ij}^{\otimes}(\hat{\mathbf{k}})=  \frac{1}{\sqrt{2}} [ e_i^1 (\hat{\mathbf{k}})  e_j^2 (\hat{\mathbf{k}})+   e_i^2 (\hat{\mathbf{k}})  e_j^1 (\hat{\mathbf{k}})] .
 \label{eq:polten}
\end{equation}

With these definitions the polarization tensors are normalized as $e_{ij}^A (\hat{\mathbf{k}})\,e^{B,ij} (\hat{\mathbf{k}})= \delta^{AB}$. The polarization tensors depend on the direction $\hat{\mathbf{k}}$ since the frame $( \hat{\mathbf{k}}, \mathbf{e}_1, \mathbf{e}_2)$, which reveals the 2 physical d.o.f.\ of the spin-2 field in eq.~(\ref{eq:fourexp}), depends on $\hat{\mathbf{k}}$. In this frame the spin-2 field has non zero components only in the plane $(\mathbf{e}_1, \mathbf{e}_2)$ and, for fixed $\hat{\mathbf{k}}$, we can always choose the basis $(e_1 (\hat{\mathbf{k}}),e_2 (\hat{\mathbf{k}}))$ in which

\begin{equation}(e_{ab }^{\oplus})= \frac{1}{\sqrt{2}}\begin{pmatrix}
1 & 0 \\
0&-1
\end{pmatrix} , \quad  (e_{ab }^{\otimes})= \frac{1}{\sqrt{2}}\begin{pmatrix}
0& 1 \\
1&0
\end{pmatrix} .\end{equation}

We can also write the tensor perturbation in the helicity basis $ \mathbf{e_{\pm}}= \frac{1}{\sqrt2} (\mathbf{e}_1 \mp i \mathbf{e}_2)  $, as

\begin{equation}  \mathbf{h}^H_{\mathbf{k}}= \frac{1}{\sqrt{2}} \begin{pmatrix}
  h_{\mathbf{k}}^{\oplus} - i  h_{\mathbf{k}}^{\otimes} &0\\
 0 &  h_{\mathbf{k}}^{\oplus} + i h_{\mathbf{k}}^{\otimes}
 \end{pmatrix} = \begin{pmatrix}
 h_{\mathbf{k}}^+ & 0 \\
0& h_{\mathbf{k}}^-
\end{pmatrix}= h_{\mathbf{k}}^+ e_{ij}^+ +  h_{\mathbf{k}}^- e_{ij}^- \;, \end{equation}
where we have defined $e_{ab}^{\pm} \equiv e_a^\pm e_b^\pm$. We have also set $ h_{\mathbf{k}}^{\pm} \equiv \frac{1}{\sqrt{2}}  ( h_{\mathbf{k}}^{\oplus} \mp i  h_{\mathbf{k}}^{\otimes} )$ and we shall work with these from now on.\footnote{Working with $ h_{\mathbf{k}}^{\pm}$ instead of the usual  $ h_{\mathbf{k}}^{\oplus,\otimes}$ is more convenient and, for parity invariant perturbations, does not make any difference since all the power spectra are equal:\\ $\braket{ h_{\mathbf{k}}^+  h_{\mathbf{k}}^{+*}}= \frac{1}{2} (\braket{h_{\mathbf{k}}^{\oplus} h_{\mathbf{k}}^{\oplus *}}+\braket{h_{\mathbf{k}}^{\otimes} h_{\mathbf{k}}^{\otimes *}})= \braket{h_{\mathbf{k}}^{\oplus} h_{\mathbf{k}}^{\oplus*}} =\braket{ h_{\mathbf{k}}^-  h_{\mathbf{k}}^{-*}}$ }
 Let us now rewrite the Fourier components of the GWs in the spherical basis of eq.~(\ref{eq:pertmetrconf}). In other words we perform a rotation with Euler angles ($\alpha,\beta,\gamma$) to rotate $( \hat{\mathbf{k}}, \mathbf{e}_1, \mathbf{e}_2)$ into $({\mathbf{n}},\mathbf{e}_{\theta},\mathbf{e}_{\phi})$. After this operation we can write (for more details see Appendix~\ref{appA}) 
 
 \begin{equation} \begin{split}
& h_{\mathbf{k}}^{rr} =  \frac{1}{2}  \sin^2 \beta \, (  e^{2 i \gamma} \: h_{\mathbf{k}}^+ +  e^{-2 i \gamma} \: h_{\mathbf{k}}^-) \;,\\
& h_{\mathbf{k}}^{r\pm} \equiv \frac{1}{\sqrt2} ( h_{\mathbf{k}}^{r\theta} \mp i  h_{\mathbf{k}}^{r\phi})= \frac{\sin\beta \left((\cos \beta \mp 1) e^{2 i \gamma }  \; h_{\mathbf{k}}^+  + (\cos \beta  \pm 1)e^{-2 i \gamma } \; h_{\mathbf{k}}^-  \right)}{2 \sqrt{2}} \; e^{i \alpha} .
\end{split}
\label{eq:hsphetomod}
\end{equation}

With this and eq.~(\ref{eq:defang}) we can now write $\alpha_\pm$ in Fourier space. We then  solve eq.~(\ref{eq:system2}) to find expressions for the lensing potentials $\psi$ and $\varpi$ in Fourier space in terms of $h_{\mathbf{k}}^{rr}$ and $ h_{\mathbf{k}}^{r\pm}$. We want to compute their angular power spectra in terms of  the power spectrum of$h_{\mathbf{k}}^\pm$ given  by

\begin{equation}
\braket{h^+_{\mathbf{k}}(\chi) h^{+*}_{\mathbf{k'}}(\chi')} = \braket{h^-_{\mathbf{k}}(\chi) h^{-*}_{\mathbf{k'}}(\chi')}= (2\pi)^3 \delta^3 (\mathbf{k-k'}) P_h(k;\chi,\chi') .
\end{equation}

Since we observe the displacement vector on the celestial sphere the natural expansion for the lensing potentials is in terms of spherical harmonics

\begin{equation} \begin{split}
& \psi ({\mathbf{ n}}) = \sum_{\ell m} a_{\ell m} Y_{\ell m} ({\mathbf{ n}}) ,\\
& \varpi ({\mathbf{ n}}) = \sum_{\ell m} b_{\ell m} Y_{\ell m} ({\mathbf{ n}}) . \\
\end{split}\label{eq:potdec}\end{equation}

In order to find the harmonic coefficients, the generalized addition relations for spherical harmonics are required: for a rotation from $(\theta_k,\phi_k)$ to $(\theta,\phi)$ with Euler angles $(\alpha,\beta,\gamma)$, such as the one performed before, we have~\cite{Ruth_book}

\begin{equation}
\sqrt{\frac{4 \pi}{2\ell+1}} \, \, \sum_{m'} {}_s Y_{\ell m'}(\theta_k,\phi_k) \, \, {}_{m} Y_{\ell m'}^* (\theta,\phi) = {}_s Y_{\ell m} (\beta,\alpha) e^{-i s \gamma} .
\label{eq:Eulspherical}
\end{equation}

As it should, this exactly eliminates the Euler angle dependence (coming from the solutions of eq.~(\ref{eq:system2})) in the expressions for $\psi$ and $\varpi$. The dependence on $(\theta_k,\phi_k)$ can be integrated out performing the angular integral in k-space and we only have to recast the dependence on the direction of observation  $(\theta,\phi)$ into the $ Y_{\ell m} (\theta,\phi)$ in order to read out the harmonics expansion coefficients $a_{\ell m}$ and $b_{\ell m}$. For a statistically isotropic field $\A({\mathbf{n}})$ the angular power spectrum $C_{\ell m}^\A$ is given  by $\braket{\A_{\ell m} \A_{\ell'm'}^*}= \delta_{\ell\ell'} \delta_{mm'} C_\ell^\A$ where the $\A_{\ell m}$s are its harmonics coefficients. For the E-mode and B-mode lensing potentials we obtain
(see~\cite{Vittorio-thesis} for more details)

\begin{equation}
\begin{split}
C_\ell^\psi =& 2 \pi \frac{(\ell+2)(\ell-1)}{\ell(\ell+1)} \int \frac{\td k}{k} \Delta_h^2 (k) \\
&\times \left| \int_{0}^{\chi_*} \td \chi \, \,  T_h(k,\chi) \left( \frac{\ell+1}{2}\left( \ell \frac{\chi-\chi_*}{\chi_*}+2 \right) \frac{j_\ell(k \chi)}{k^2 \chi^3} -\frac{j_{\ell+1}(k\chi)}{k \chi^2} \right) \right|^2 
\label{eq:scalarpotential}
\end{split}
\end{equation}
and

\begin{equation}
C_\ell^\varpi = \pi \frac{(\ell+2)(\ell-1)}{\ell(\ell+1)} \int \frac{\td k}{k} \, \Delta_h^2 (k) \left| \int_{0}^{\chi_*} \td \chi \, \, T_h(k,\chi) \frac{j_\ell(k\chi)}{k \chi^2} \right|^2 .
\label{eq:Clomega}
\end{equation}

Here we have written the power spectrum $P_h(k;\chi,\chi')$ in the form 

\begin{equation}
\frac{k^3}{2\pi^2}P_h(k;\chi,\chi')= \Delta_h^2 (k) T_h(k,\chi) T_h(k,\chi') , 
\label{eq:tspectrum}
\end{equation}
where $\Delta_h^2 (k)$ is the dimensionless primordial power spectrum and $ T_h(k,\chi)$ is the tensor transfer function. We show numerical results for the E- and B-mode lensing potentials in Section~\ref{s:num}, Figs.~\ref{fig:plot1} and \ref{fig:plot2}.

The case $\ell = 2$ (quadrupole) of Eq.~(\ref{eq:scalarpotential}) is peculiar and needs to be discussed carefully. On the one hand, one may note that the $k$-integral is infrared divergent for any
initial power spectrum without a blue tilt. On the other hand, even if no infrared divergence was present, the limit $\chi_* \rightarrow 0$ generally yields a non-zero result. As will become clear from the
discussion in Section \ref{s:ani}, the effect of long modes locally looks like an anisotropic Bianchi I model, i.e.\ a spacetime in which comoving rulers pointing along different axes will expand at
different rates. From the point of view of the observer it makes sense to redefine the coordinates such that this effect vanishes locally, i.e.\ to choose identical rulers in all directions
at the time of observation. Due to the evolution of the tensor perturbations the rulers may not remain identical for a long time, but at least observers choose rulers in the different directions such that no lensing occurs
in the limit of $\chi_* \rightarrow 0$. To achieve this mathematically, we simply have to subtract the limiting expression for $\chi_* \rightarrow 0$ from the full expression given in Eq.~(\ref{eq:scalarpotential}). Taking the limit
inside the $k$-integral also regulates the infrared divergence. We argue that this is the correct prescription for the observational procedure: the observables are only affected by the presence
of tensor perturbations which vary over the observed patch of spacetime, while a ``constant perturbation'' should be absorbed into the choice of coordinates\footnote{It is necessary to take care of this subtlety only in tensor perturbations lensing since  for lensing from vector perturbations it affects only the dipole and for scalar lensing it affects the monopole. These gauge dependent multipoles are usually not considered.}. The regularized expression for the quadrupole is then

\begin{equation*}
C_2^{\psi\,\text{(reg)}}=\frac{4 \pi}{3} \int \frac{\td k}{k} \, \Delta_h^2 (k) \left| \int_{0}^{\chi_*} \td \chi \, \,\left[ \frac{3}{\chi_*} \left( \frac{j_2(k \chi)}{k^2 \chi^2} T_h (k,\chi)- \frac{T_h(k,0)}{15} \right) - \frac{j_3 (k \chi)}{k \chi^2} T_h(k,\chi) \right] \right|^2.
\end{equation*}

This regularization procedure has not been discussed previously, but when computing the lensing potentials by the total angular momentum method the subtraction which we have introduced here by hand appears as a boundary term, see~\cite{Yamauchi:2013fra}.

\subsection{The lensed CMB  power spectrum}

\newcommand{\ti}{\tilde}

Having computed the power spectra of the the two lensing potentials from tensor perturbations, we can now determine how gravitational lensing affects the shape of the CMB temperature angular power spectrum $C_\ell^{\Theta}$, where $\Theta ({\mathbf{n}}) \equiv (T({\mathbf{n}})-\braket{T})/\braket{T}$ is the temperature anisotropy field (see~\cite{Ruth_book}). In other words we want to compute the lensed CMB power spectrum $\ti C_\ell^\Theta$. The full-sky formalism developed in~\cite{Hu:2000ee} is preferable to the approach of~\cite{Lewis:2006fu} since the former is based on a Taylor expansion in the displacement vector $\boldsymbol{\alpha}$ which we expect to be very small in the case of lensing by tensor perturbations. We can support this claim by quickly computing the rms deflection angle for lensing by primordial gravitational waves with a tensor/scalar ratio $r=0.2$. We find

\begin{equation}
 \theta_{\text{rms}}^2=  \braket{| \nabla \psi + \nabla \times \varpi |^2} \simeq 1.2 \times 10^{-9} .
\end{equation}
to give $\theta_{\text{GW}} \simeq 7 \; \; \text{arcsec}$ which is 20 times smaller than the value for density perturbations $\theta_{\delta} \simeq 2.7 \; \; \text{arcmin}$ \cite{Lewis:2006fu}. The harmonic approach developed in~\cite{Hu:2000ee} for density perturbations is applied to tensor perturbations in~\cite{Li:2006si} and we  follow the same approach. Assuming a small deflection angle $\boldsymbol{\alpha}$ we Taylor expand $\ti \Theta ({\mathbf{n}}) = \Theta({\mathbf{n}}+ \boldsymbol{\alpha})$ to second order

\begin{equation}
\ti \Theta ({\mathbf{ n}}) = \Theta({\mathbf{ n}}+ \boldsymbol{\alpha}) \simeq \Theta ({\mathbf{ n}}) + \nabla^a \Theta \, \alpha_a + \frac{1}{2} \nabla^b \nabla^a \Theta \, \alpha_a \alpha_b + ...
\label{eq:taylor}
\end{equation}

Recalling the angular decompositions of eq.~(\ref{eq:potdec}) and defining $\theta_{\ell m}$ as the harmonic coefficients of the temperature anisotropies field $ \Theta ({\mathbf{n}}) = \sum_{\ell m} \theta_{\ell m} Y_{\ell m}({\mathbf{n}})$, eq.~(\ref{eq:taylor}) yields

\begin{equation}
\begin{aligned}
& \ti \theta_{\ell m} =\, \theta_{\ell m} + \int d\Omega \left(\nabla^a \Theta \, \alpha_a + \frac{1}{2} \nabla^b \nabla^a \Theta \, \alpha_a \alpha_b  \right) Y_{\ell m}^*\\
& =\, \theta_{\ell m} + \sum_{\ell_1 m_1 \ell_2 m_2} \theta_{\ell_1 m_1} \left( a_{\ell_2 m_2} I^{(a)}_{\ell m\ell_1 m_1 \ell_2 m_2} + b_{\ell_2 m_2} I^{(b)}_{\ell m\ell_1 m_1 \ell_2 m_2} \right) \\
& +\frac{1}{2} \sum_{\substack{\ell_1  \ell_2 \ell_3 \\ m_1 m_2 m_3}} \theta_{\ell_1 m_1}  \left(  a_{\ell_2 m_2} a_{\ell_3 m_3}^* K^{(a)}_{\ell m\ell_1 m_1 \ell_2 m_2 \ell_3 m_3} + b_{\ell_2 m_2} b_{\ell_3 m_3}^* K^{(b)}_{\ell m \ell_1 m_1 \ell_2 m_2 \ell_3 m_3}  \right), 
\label{eq:angmomapp}
\end{aligned}
\end{equation}

where for the last equality we have used $\alpha_a = \nabla_a \psi + \varepsilon^b {}_a \nabla_b \varpi$ and we have defined the following integrals

\begin{equation}\begin{aligned}
& I^{(a)}_{\ell m\ell_1 m_1 \ell_2 m_2}=\int d\Omega \left( \nabla^a Y_{\ell_1 m_1}\right) Y_{\ell m}^* \nabla_a Y_{\ell_2 m_2} , \\
& I^{(b)}_{\ell m\ell_1 m_1 \ell_2 m_2}=\int d\Omega \left( \nabla^a Y_{\ell_1 m_1} \right)Y_{\ell m}^* \varepsilon_a^b \nabla_b Y_{\ell_2 m_2}   ,\\
& K^{(a)}_{\ell m\ell_1 m_1 \ell_2 m_2 \ell_3 m_3}=\int d\Omega \left( \nabla^b \nabla^a Y_{\ell_1 m_1}\right) Y_{\ell m}^* \left(\nabla_a Y_{\ell_2 m_2}\right) \nabla_b Y_{\ell_3 m_3}^*  , \\
&  K^{(b)}_{\ell m\ell_1 m_1 \ell_2 m_2 \ell_3 m_3}=\int d\Omega \left( \nabla^b \nabla^a Y_{\ell_1 m_1}  \right)Y_{\ell m}^* \varepsilon_a^c \left(\nabla_c Y_{\ell_2 m_2} \right) \varepsilon_b^d \nabla_d Y_{\ell_3 m_3}^* \, .
\end{aligned}
\end{equation}

Our aim is to compute $\braket{\ti \theta_{\ell m} \ti \theta_{\ell'm'}^*}=\delta_{\ell\ell'}\delta_{mm'}C_\ell^{\ti \Theta}$. For this we substitute eq.~(\ref{eq:angmomapp}) into the two-point angular correlation function.  After some algebra this leads to

\begin{equation}
C_\ell^{\ti \Theta}= C_\ell^{\Theta} + \sum_{\ell_1 \ell_2} C_{\ell_1}^{\Theta} \left( C_{\ell_2}^{\psi} \, \Pi^{(1a)}_{\ell \ell_1 \ell_2} + C_{\ell_2}^{\varpi} \, \Pi^{(1b)}_{\ell \ell_1 \ell_2} \right) +\frac{1}{2} C_\ell^{\Theta} \sum_{\ell_1} \left( C_{\ell_1}^{\psi} \, \Pi^{(2a)}_{\ell \ell_1}+ C_{\ell_1}^{\varpi} \, \Pi^{(2b)}_{\ell \ell_1} \right) ,
\end{equation}
where

\begin{equation}\begin{aligned}
& \Pi^{(1a)}_{\ell \ell_1 \ell_2}= \sum_{m_1 m_2} |I^{(a)}|^2 , \qquad  \Pi^{(1b)}_{\ell \ell_1 \ell_2}=\sum_{m_1 m_2} |I^{(b)}|^2 \\
&\Pi^{(2a)}_{\ell \ell_1}= \sum_{m_1} \left( K^{(a)} + K^{(a)*} \right)  , \qquad \Pi^{(2b)}_{\ell \ell_1}=\sum_{m_1} \left( K^{(b)} + K^{(b)*} \right) .
\end{aligned}
\label{eq:Pi}
\end{equation}

Where, for simplicity, we have suppressed the indices in the summands $I^{(a,b)}$ and $K^{(a,b)}$ and only retained the indices on which the result depends on the lhs.
To obtain these expressions we made use of the fact that $\theta_{\ell m},~a_{\ell m}$ and $b_{\ell m}$ are uncorrelated Gaussian random variables with zero mean. This implies that the bispectrum vanishes. We have also neglected the trispectrum of these variables since it is of second order and of course any other higher order polyspectra are set to zero. To compute the integrals in eq.~(\ref{eq:Pi}) we use Gaunt's formula for the integration of a product of three $_s\!Y_{\ell m}$'s. We also use $\sqrt{2}\nabla_+ =-\ssd$ and $\sqrt{2}\nabla_- =-\ssu$ in order to write $\nabla_aY_{\ell m}$ in terms of $_{\pm1}\!Y_{\ell m}$ and  integration by parts to reconstruct $\De Y_{\ell m}=-\ell(\ell+1)Y_{\ell m}$.  With this we can express the integrals in terms of the Wigner-3j symbols
which are defined by
\begin{equation}\begin{aligned}
\int d\Omega \left( {}_{s_1}Y_{\ell_1 m_1}^* \right)\left( {}_{s_2}Y_{\ell_2 m_2} \right)
\left( {}_{s_3}Y_{\ell_3 m_3} \right)=& (-1)^{m_1+s_1} \sqrt{\frac{(2\ell_1+1)(2\ell_2+1)(2\ell_3+1)}{4 \pi}}\\
& \times \begin{pmatrix}
\ell_1 & \ell_2 & \ell_3 \\
s_1& -s_2 & -s_3 \end{pmatrix}  \begin{pmatrix}
\ell_1 & \ell_2 & \ell_3 \\
-m_1& m_2 & m_3 \end{pmatrix}.
\end{aligned}
\end{equation}
We also use the identity
\begin{equation*}
\sum_{m_1 m_2}  \begin{pmatrix}
\ell_1 & \ell_2 & \ell_3 \\
m_1& m_2 & m_3 \end{pmatrix}  \begin{pmatrix}
\ell_1 & \ell_2 & \ell_3 \\
m_1& m_2 & m_3 \end{pmatrix} = \frac{1}{2\ell_3+1}  \;.
\end{equation*}

In the end we find the relatively simple expressions

\begin{equation}\begin{aligned}
& \Pi^{(1a)}_{\ell \ell_1 \ell_2}= \frac{1}{16 \pi} (2\ell+1)(2\ell_1+1)(2\ell_2+1) \, \left(  [\ell(\ell+1)-\ell_1(\ell_1+1)-\ell_2(\ell_2+1)]   
\begin{pmatrix}
\ell & \ell_1 & \ell_2 \\
0& 0 & 0 \end{pmatrix} \right)^2    , \\
& \Pi^{(1b)}_{\ell \ell_1 \ell_2}= \frac{1}{16 \pi} (2\ell+1)(2\ell_1+1)(2\ell_2+1)(\ell_1(\ell_1+1)\ell_2(\ell_2+1)) \, \left(  [1-(-1)^{\ell+\ell_1+\ell_2}]   
\begin{pmatrix}
\ell & \ell_1 & \ell_2 \\
0& -1& 1 \end{pmatrix} \right)^2  , \\
&\Pi^{(2a)}_{\ell \ell_1}= \Pi^{(2b)}_{\ell \ell_1} = -\ell(\ell+1)\ell_1(\ell_1+1) \frac{2\ell_1+1}{4 \pi} .
\end{aligned}
\label{eq:Pi2}
\end{equation}

Finally, combining all the expressions, we arrive at 

\begin{equation}\begin{aligned}
C_\ell^{\ti \Theta}=& \,C_\ell^\Theta + \sum_{\ell_1 \ell_2} \frac{C_{\ell_1}^{\Theta}}{2\ell+1} \left( C_{\ell_2}^{\psi} \, \Pi^{(1a)}_{\ell \ell_1 \ell_2} + C_{\ell_2}^\varpi \, \Pi^{(1b)}_{\ell \ell_1 \ell_2} \right) \\
& -\ell(\ell+1) C_\ell^\Theta  \sum_{\ell_1} \ell_1 (\ell_1+1) \frac{2\ell_1+1}{8 \pi} \left(C_{\ell_1}^\psi + C_{\ell_1}^{\varpi} \right) ,  \label{eq:CMBlens}
\end{aligned}
\end{equation}
which is the first order lensed CMB temperature power spectrum as a function of the unlensed CMB spectrum and the lensing potentials. One can now go on and compute the lensing of CMB polarization. But due to the smallness of the polarization signal this will be even smaller than the temperature signal which we compute here and of which we shall show in the next section that it is smaller than cosmic variance in all circumstances.  

\section{Numerical results}\label{s:num}

We now evaluate numerically the angular power spectrum of the lensing E-mode potential $\psi$ and the B-mode potential $\varpi$ for tensor perturbations that we derived in~\ref{ss:1.2}. There are two types of tensor perturbations relevant for cosmology: primordial gravitational waves from inflation and those induced at second order by scalar perturbations. We treat both of these perturbations in a $\Lambda$CDM scenario.\\ 

Inflationary cosmology predicts the generation of primordial gravitational waves with a nearly scale invariant spectrum that we can parametrize, following eq.~(\ref{eq:tspectrum}), by

\begin{equation}
\Delta_h^2 (k) = \frac{k^3}{2 \pi^2} P_h^{(i)} (k) = r A_s \left(  \frac{k}{k_*} \right)^{n_t} ,
\end{equation}
where $A_s= 2.21 \times 10^{-9}$ is the amplitude of the scalar perturbation spectrum at the fiducial scale $k_*= 0.05  \; \text{Mpc}^{-1}$, $n_t \simeq 0$ is the tensor spectral index and $r$ is the tensor to scalar ratio. $A_s$ has been measured accurately in the CMB~\cite{Ade:2015xua}; primordial gravitational waves have  not been observed so far, but are limited~\cite{Ade:2015tva} to $r\lesssim 0.1$. We choose the typical inflationary value $n_t \simeq -0.013$ and set $r=0.1$.\footnote{For the value of $n_t$ were assume the second order consistency relation which yields $n_t=-\frac{r}{8}\left(2-\frac{r}{8}-n_s \right)$} The transfer function in matter or radiation domination is an analytic function of $x= k \chi$ but, if we want to include the effect of $\Lambda$ domination, we have to solve  numerically the evolution equation

\begin{equation}
\ddot{h}_k +2 \mathcal{H} \, \dot{h}_k+k^2 h_k=0 .
\label{eq:primEv}
\end{equation}

As mentioned before, at second order in perturbation theory the scalar spectrum sources the generation of secondary, or scalar-induced, tensor modes. This, contrary to the primordial gravitational waves, is an unavoidable effect and does not depend on the inflationary model or its tuning. Due to the presence of scalar perturbations, the evolution equation for tensor modes is modified to

\begin{equation}
\ddot{h}_k +2 \mathcal{H} \, \dot{h}_k+k^2 h_k= S(k,\chi) ,
\end{equation}
where $S(k,\chi)$ is a source term. At second order $S(k,\chi)$  is a convolution of two first-order scalar perturbations at different wave numbers~\cite{Ananda:2006af,Baumann:2007zm}. The initial value and the time evolution of the first-order scalar spectrum necessary for the computation of the scalar-induced tensor spectrum $P_h^{(2)}(k,\chi)$ are obtained from the publicly available Boltzmann code CLASS~\cite{Lesgourgues:2011re}, while the primordial tensor evolution can be obtained by numerical integration of eq.~(\ref{eq:primEv}). Furthermore we take advantage of a recent work, by two of us, on relativistic N-body simulations \cite{Adamek:2015eda}. This allows us to obtain the fully nonperturbative induced tensor spectrum $P_h^{(\mathrm{full})}(k,\chi)$ which, as opposed to $P_h^{(2)}(k,\chi)$, contains the effect of fully nonlinear small scale structure. To simplify the numerical computation we use Limber's approximation

\begin{equation}\label{eq:limber}
\int k^2 \td k \, j_\ell(k \chi) j_\ell (k \chi') \simeq \frac{\pi}{ 2 \chi^2} \delta(\chi-\chi') \;, 
\end{equation}
to reduce the dimensionality of the integrals in eq.~(\ref{eq:scalarpotential}) and~(\ref{eq:Clomega}). We obtain

\begin{equation}
\begin{split}
C_\ell^\psi \simeq \, & 4 \pi^2 \frac{(\ell+2)(\ell-1)}{\ell(\ell+1)} \int_{0}^{\chi_*} \frac{\td \chi}{\chi} \; \Bigg\{ \frac{1}{(\ell+1)^2(2\ell+3)^2} \left[ \Delta_h^2 (k)  \, \, T_h^2(k,\chi) \; \right]_{k=(\ell+1)/\chi}    \\
& \frac{\ell+1}{2 \ell^4(2\ell+1)^2} \left( \frac{\ell+1}{2\ell} \left(\ell \frac{\chi -\chi_*}{\chi_*}+2 \right)^2 -2 \left(\ell \frac{\chi -\chi_*}{\chi_*}+2 \right)^2 \right) \left[ \Delta_h^2 (k)  \, \, T_h^2(k,\chi) \; \right]_{k=\ell/\chi} \Bigg\} ,
\end{split}
\label{eq:Clpsilimber}
\end{equation}

\begin{equation}
C_\ell^\varpi \simeq \frac{\pi^2}{2 \ell^5} \frac{(\ell+2)(\ell-1)}{\ell(\ell+1)} \int_{0}^{\chi_*} \frac{\td \chi}{\chi} \; \left[ \Delta_h^2 (k)  \, \, T_h^2(k,\chi) \; \right]_{k=\ell/\chi}  .
\label{eq:Clomegalimber}
\end{equation}

While we have used the Limber approximation (\ref{eq:limber}) for (\ref{eq:Clomegalimber}) we have also used it for integrals of the type $j_\ell(k\chi)j_{\ell+1}(k\chi)$ in Eq.~(\ref{eq:Clpsilimber}). Even though it is not derived for this case, we have checked numerically that, for sufficiently large $\ell$, it is a good approximation also in this case.
The Limber approximation is usually reasonable for $\ell\gtrsim 10$ and it improves as $\ell$ increases. However this is only true if the function integrated with the spherical Bessels $j_\ell(k \chi)$ is slowly varying. This causes no problem for $P_h^{(2)}(k,\chi)$ and $P_h^{(\text{full})}(k,\chi)$, but the transfer function for primordial gravitational waves oscillates very rapidly at small scales and Limber's approximation gets worse as $\ell$ increases so that we are left with no choice but to compute the double integrals of eq.~(\ref{eq:scalarpotential}) and~(\ref{eq:Clomega}). \\

    \begin{figure}[h]	
\centering
    	\includegraphics[scale=0.37]{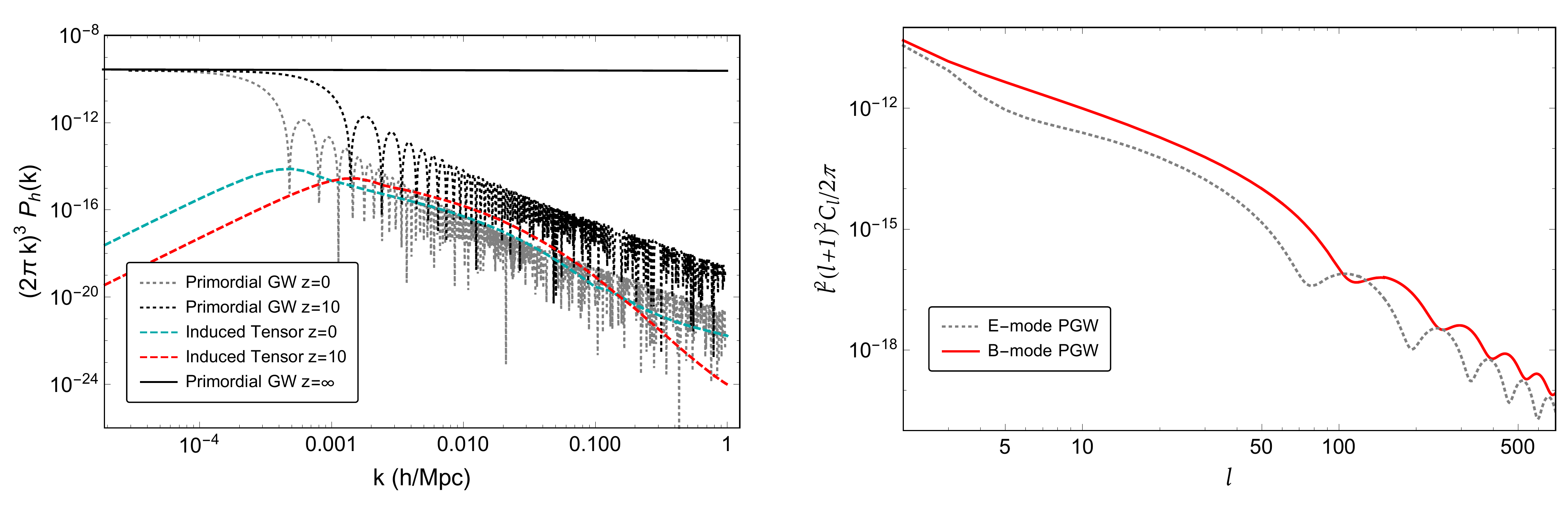}
	\caption{\emph{Left}: The primordial [gray, black] and the scalar-induced [cyan, red] power spectra for $z=0$ and $z=10$. \emph{Right}: The B-mode [red] and the E-mode [dotted] lensing potentials angular spectra induced by primordial gravitational waves.}
   	 \label{fig:plot1}
\end{figure}

    \begin{figure}[h]	
\centering
    	\includegraphics[scale=0.37]{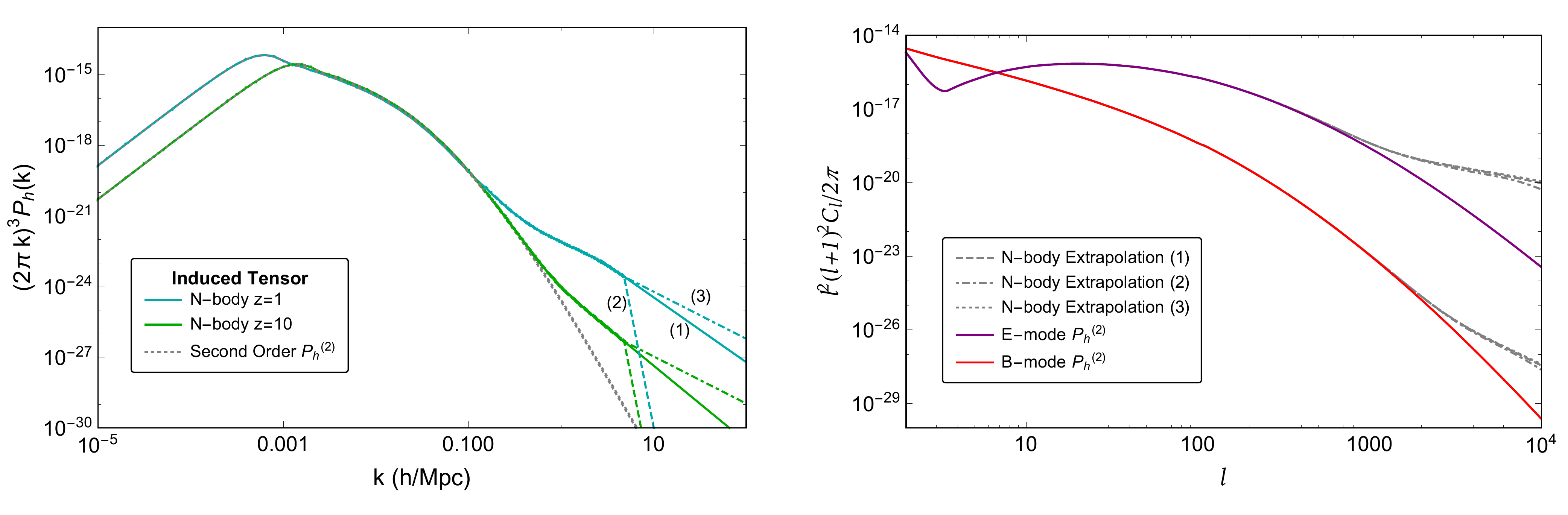}
	\caption{\emph{Left}: The scalar-induced tensor power spectrum for $z=1$ [cyan] and $z=10$ [green]. Different types of small scale extrapolations are shown: power law extrapolation (1) [solid], exponential cutoff (2) [dashed] and $\propto k^{-2}$ behavior (3) [dot-dashed]. The second order spectrum $P_h^{(2)}$ is also shown [gray, dotted] for the two redshifts. \emph{Right}: The E-mode [purple] and B-mode [red] lensing potentials of induced tensors. The spectra from the second order calculation are compared to the fully nonperturbative results [gray].}
   	 \label{fig:plot2}
\end{figure}

In Fig.~\ref{fig:plot1} we plot the primordial tensor power spectrum and its E-mode and B-mode lensing spectra for a cosmology with $\Omega_b = 0.05$ , 
$\Omega_{\text{cdm}} = 0.26$, $\Omega_{\Lambda} = 0.69$, $h=0.68$ and $n_s = 0.96$. Primordial tensor modes are constant outside the horizon ($kt<1$) and they start to oscillate and decay like $1/a$ once inside the horizon ($kt<1$).  \\

In Fig.~\ref{fig:plot2} the induced tensor power spectrum is shown, together with its lensing spectra. The tilt of the second order spectrum outside the horizon, for $k \rightarrow 0$, is given by $k^3 P_h^{(2)} (k) \propto k^3$ and one can also see the enhancement of power due to non linearities at late time and small scales. The numerical data of \cite{Adamek:2015eda} extends to $k \simeq \, 2\text{h/Mpc}$ and we present three types of small scale extrapolations to check the robustness of the nonlinear effects in the lensing signals. We find that, for the multipoles shown, the effect is dominated by the scales resolved in the simulations, and therefore its amplitude is nearly independent of the extrapolation to smaller scales. We also show the second order spectrum $P_h^{(2)}(k,\chi)$.

Let us discuss this somewhat more precisely: The Limber's approximation implies that the main contribution to the lensing potential mode $\ell$ comes from all values of $k$ with $k \simeq \ell/\chi$ for $\chi\in [0,\chi_*]$. But since the induced tensor power spectrum is decaying with $k$ for $k>k_{\rm eq}\simeq 0.001h/$Mpc, see Fig~\ref{fig:plot2}, the  dominant contribution for $\ell\gtrsim 50$ comes from $k_*(\ell)=\ell/\chi_* \simeq \ell/t_0 \simeq (\ell/10^4)h/$Mpc. Hence for $\ell \lesssim 10^4$ the dominant contribution comes from within the numerical simulation which go up to $k \simeq \, 2h/\text{Mpc}$. This explains the weak dependence of the lensing power spectrum  below $\ell \simeq 10^4$ on the very small scales extrapolations of the induced tensor power spectrum. It is interesting to note that nonlinearities enhance the tensor lensing potential on small scales, $\ell>1000$ by up to several orders of magnitude.
We should also point out that we neglect the decaying gravitational wave background which has been
induced during the radiation era and which, at second order, becomes relevant at the smallest scales (see \cite{Baumann:2007zm} for a discussion of this effect). \\

\begin{figure}[h]	
\centering
    	\includegraphics[scale=0.37]{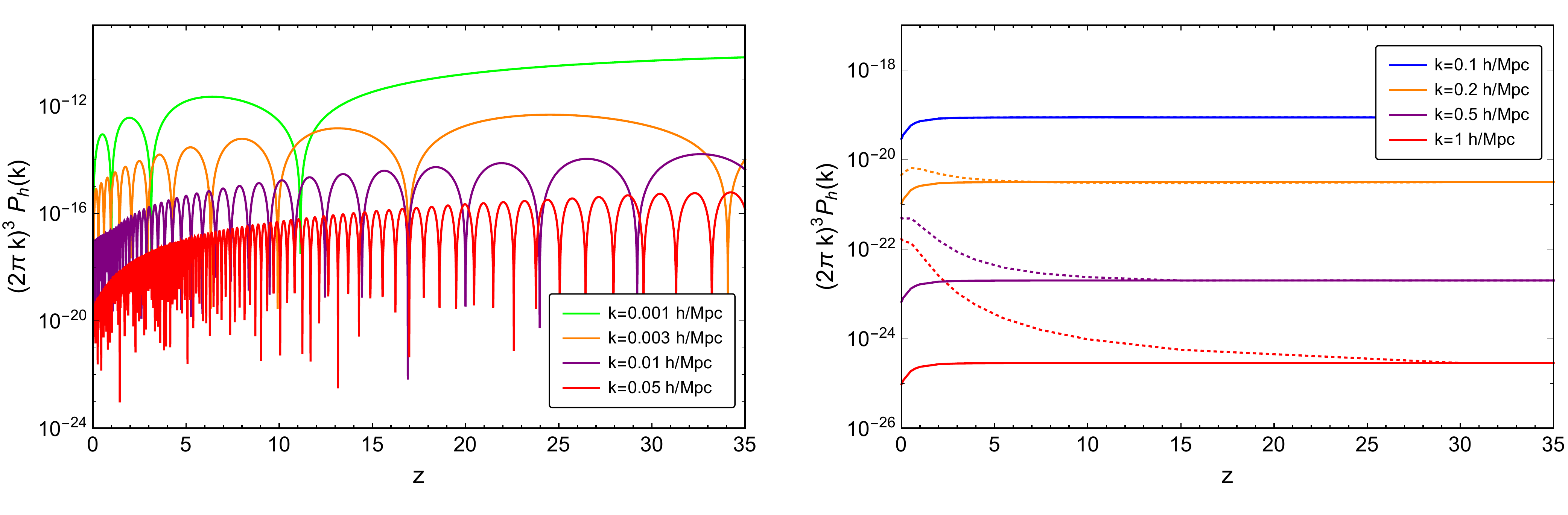}
	\caption{\emph{Left}: The primordial gravitational wave spectral power for different wavenumbers as a function of $z$. \emph{Right}: The second order scalar-induced tensor spectral power [solid] for different wave numbers as a function of $z$ together with the late-time effect from nonlinearities [dashed].}
   	 \label{fig:plot3}
\end{figure}

In Fig.~\ref{fig:plot3} we plot the time evolution of the primordial and scalar-induced tensor modes. We can see that at small scales and late times non linearities enhance the power of the second order spectrum. \\

\begin{figure}[h]	
\centering
    	\includegraphics[scale=0.42]{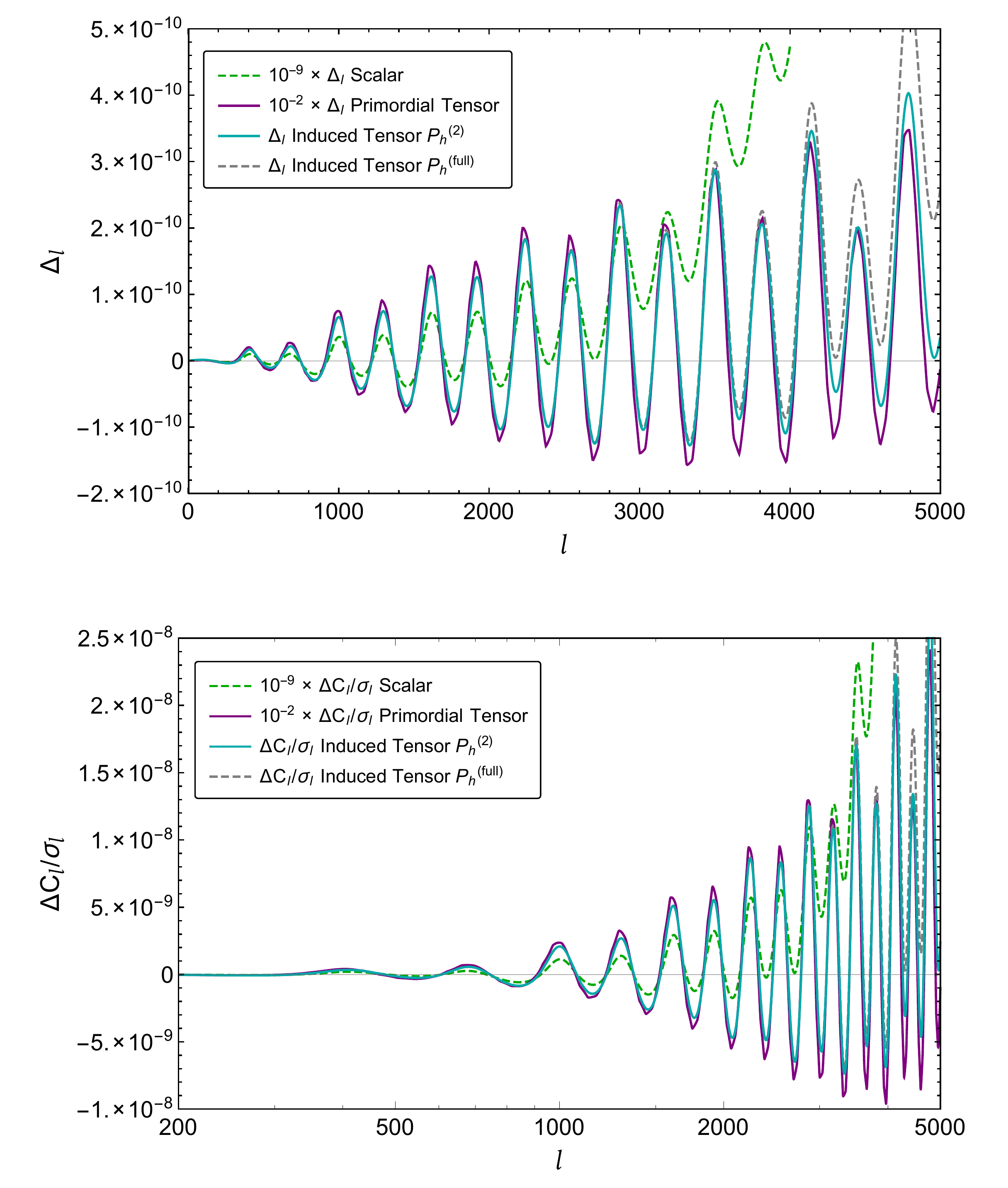}
	\caption{\emph{Top}: The CMB angular power spectrum fractional difference $\Delta_\ell$ for scalar (multiplied by $10^{-9}$) [green, dashed] and tensor type lensing both for PGW (multiplied by $10^{-2}$) [purple] and scalar-induced tensor [cyan] \emph{Bottom}: The relative difference $(C_\ell^{\ti \Theta}-C_\ell^{\Theta})/\si_\ell$, where $\si_\ell^2=2C_\ell^2/(2\ell+1)$ is the error from cosmic variance.}
   	 \label{fig:plot4}
\end{figure}

In Fig.~\ref{fig:plot4} we consider the effect of Spin-2 perturbation on the CMB temperature spectrum. We use the result of eq.~(\ref{eq:CMBlens}) to evaluate the fractional difference \[ \Delta_\ell=\frac{C_\ell^{\ti \Theta}-C_\ell^{\Theta}}{C_\ell^{\Theta}}\] between the unlensed and the lensed spectrum both for primordial and scalar-induced tensor perturbation.

It is important to note that the scalar-induced tensor power spectrum is not a gravitational wave in the usual sense. Most of its amplitude is frozen in as a scale dependent anisotropy of spacetime and is not oscillating. Such a spectrum of sourced tensor perturbations would actually not show up in  interferometric experiments like LIGO~\cite{ligo:Online} or eLISA~\cite{elisa:Online} which measure the oscillations of the spacetime geometry as a function of time.  However, it can in principle be observed via its lensing effect.  Here we have studied the effect of tensor-lensing on the CMB. The final result given in eq.~(\ref{eq:CMBlens}) is shown in Fig.~\ref{fig:plot4}. We have found that the difference between the tensor-lensed and the unlensed CMB angular power spectrum is smaller than cosmic variance on all scales. The reason for this is twofold. First the gravitational wave deflection angle is about 20 times smaller than the one from scalar perturbations. Secondly, the gravitational wave signal dominates on large scales where lensing deflections of the CMB only have a very small effect. The induced tensor modes might however be measurable with weak lensing or tidal effects on galaxy alignment measurements as proposed e.g. in~\cite{Schmidt:2012nw,Schmidt:2013gwa}, or with lensing reconstruction and delensing techniques as proposed in~\cite{Namikawa:2014lla}.

\FloatBarrier

\section{Application to anisotropic cosmology}\label{s:ani}

In this section we want to present a  different application of our formalism. We consider a homogeneous but anisotropic universe of Bianchi I class, with line element
\begin{equation}
 \td s^2= a^2(t)[-\td t^2 + e^{2\beta_i(t)} \delta_{ij}\td x^i \td x^j] \;,
\end{equation}
where $a(t)$ is chosen such that $\sum_i \beta_i(t) = 0$ to ensure that the comoving volume evolves like $a^3$. The $\beta_i(t)$ give rise to anisotropic expansion and, evidently, taking $\beta_i(t) \rightarrow 0$
restores isotropy and gives the flat Friedmann model. If the $\beta_i(t)$ are small, i.e.\ if $\beta_i(t) \ll 1$, we can interpret this metric as a perturbed
Friedmann model with $h_{ij} = 2 \beta_i(t) \delta_{ij}$. Lensing in this type of geometry has recently been studied in \cite{Pitrou:2015iya}. The traceless tensor $h_{ij}$ can be seen as an infrared limit of the spin-2 perturbations we have discussed in the previous sections. However, the transverse condition does
not place any constraint on a spatially uniform perturbation, which means that $h_{ij}$ has three additional degrees of freedom. Indeed, we have to pick an
element of $SO(3)$ in order to specify the coordinate system in which $h_{ij}$ is diagonal (three d.o.f.) and then, in this coordinate system, give two of the $\beta_i$'s,
e.g.\ $\beta_1$ and $\beta_2$, the third $\beta_i$ being fixed by the traceless condition. As we will see, the three additional degrees of freedom arise
in the guise of a prescription of how to take the limit of infinite wavelength. Motivated by this consideration, let us examine the situation where $h_{ij}$ is given by a single mode of finite wavelength. Instead of the stochastic field  discussed in the previous sections,
we consider a gravitational wave with fixed wave number $\mathbf{k}_0$ so that  $h_{\mathbf{k}}^\pm(t) = A^\pm(t) \delta^3(\mathbf{k}-\mathbf{k}_0)$. Inserting this into the expression for the multipole coefficients of the gradient-mode lensing potential gives

\begin{equation*}\begin{aligned}
 a_{\ell m} =& i^l \sqrt{\frac{(\ell-2)!}{(\ell+2)!}} \frac{(\ell+2) (\ell-1)}{(2 \pi)^2} \int_{0}^{\chi_*} \td\chi \left( \frac{\ell+1}{2}\left( \ell \frac{\chi-\chi_*}{\chi_*}+2 \right) \frac{j_\ell(k_0 \chi)}{k_0^2 \chi^3} -\frac{j_{\ell+1}(k_0\chi)}{k_0 \chi^2} \right) \\ &\times \left({}_{-2} Y_{\ell m}(\mathbf{\hat{k}}_0) A^+(\chi) + {}_{+2} Y_{\ell m}(\mathbf{\hat{k}}_0) A^-(\chi)\right) \;,
\end{aligned}\end{equation*}
and a corresponding expression for the curl-mode potential. In the limit of infinite wavelength,  $\mathbf{k}_0\rightarrow 0$ the curl-mode potential vanishes for all $\ell$, while the gradient-mode  potential is non-zero only for $\ell=2$, where we find

\begin{equation}
 a_{2m} = - \frac{1}{10 \sqrt{6} \, \pi^2 \chi_*} \int_{0}^{\chi_*} \td\chi \left({}_{-2} Y_{2m}(\mathbf{\hat{k}}_0) A^+(\chi) + {}_{+2} Y_{2m}(\mathbf{\hat{k}}_0) A^-(\chi)\right).
\end{equation}

Note that we take the limit along some fixed direction $\mathbf{\hat{k}}_0$, and the limiting expression does depend on this choice. This is related to the issue that the polarization
tensors in eq.~(\ref{eq:fourexp}) are not well-defined at $\mathbf{k} = \mathbf{0}$. We simply define them through a limiting procedure which is, however, not unique. Let us see whether  we can obtain $h_{ij} = 2 \beta_i(t) \delta_{ij}$ by such a limiting procedure. Since the anisotropic expansion is triaxial, we use the sum of two modes whose
limits are taken along orthogonal directions. For the first mode, we choose the coordinate frame $(\mathbf{\hat{k}}_0, \mathbf{e}_1, \mathbf{e}_2) = (\mathbf{\hat{x}}_1, \mathbf{\hat{x}}_2, \mathbf{\hat{x}}_3)$ where $h_{ij} = \mathrm{diag}(2 \beta_1, 2 \beta_2, 2 \beta_3)$. We set the first mode $h_{\mathbf{k}}^\oplus = \sqrt{8} (2 \pi)^3 \delta^3(\mathbf{k}-\mathbf{k}_0) \beta_2$.
Next, we rotate our coordinate frame by $90$ degrees around the $\mathbf{e}_2 = \mathbf{\hat{x}}_3$ axis such that $(\mathbf{\hat{k}}'_0, \mathbf{e}'_1, \mathbf{e}'_2) = (\mathbf{\hat{x}}_2, -\mathbf{\hat{x}}_1, \mathbf{\hat{x}}_3)$, and set the second mode ${h_{\mathbf{k}}^\oplus}' = \sqrt{8} (2 \pi)^3 \delta^3(\mathbf{k}-\mathbf{k}'_0) \beta_1$. Now, taking the limits
$\mathbf{k}_0 \rightarrow \mathbf{0}$, $\mathbf{k}'_0 \rightarrow \mathbf{0}$, keeping the directions $\mathbf{\hat{k}}_0$, $\mathbf{\hat{k}}'_0$ fixed, we see from eq.~(\ref{eq:fourexp}) that we recover $h_{ij} = 2 \beta_i(t) \delta_{ij}$
as desired. Working out the coefficients $A^\pm$ in the helicity basis we finally arrive at

\begin{equation*}
 a_{2m} = \frac{\pi}{\sqrt{6}}\frac{8}{5 \chi_*} \int_{0}^{\chi_*} \td\chi \Big[\big({}_{-2} Y_{2m}(\mathbf{\hat{x}}_1) + {}_{+2} Y_{2m}(\mathbf{\hat{x}}_1)\big) \beta_2(\chi) + \big({}_{-2} Y_{2m}(\mathbf{\hat{x}}_2) + {}_{+2} Y_{2m}(\mathbf{\hat{x}}_2)\big) \beta_1(\chi)\Big],
\end{equation*}
explicitly

\begin{equation}\begin{aligned}
a_{20}=& -\sqrt{\frac{\pi}{5}} \, \frac{2}{\chi_*}  \int_{0}^{\chi_*} \td\chi \left( \beta_1(\chi) +\beta_2(\chi) \right), \\
 a_{2 \pm1}=& \,0, \\
 a_{2\pm2}=& \,\sqrt{\frac{\pi}{30}} \, \frac{2}{\chi_*}  \int_{0}^{\chi_*} \td\chi \left( \beta_1(\chi) -\beta_2(\chi) \right),
\end{aligned}
\end{equation}
in agreement with eqs.~(7.7) and (7.8) of \cite{Pitrou:2015iya}. As explained at the end of Section \ref{ss:1.2}, the coordinates can (and should) be chosen such that the lensing signal vanishes
in the limit $\chi_* \rightarrow 0$. In other words, one should rescale the comoving axes such that $\beta_i(\chi) \rightarrow \beta_i(\chi) - \beta_i(0)$. Equivalently, one can simply subtract
the signal which is obtained when taking $\chi_* \rightarrow 0$ (see also the comment below Eq.~(6.14) of \cite{Pitrou:2015iya}).

At the linear level, anisotropic expansion produces a quadrupolar, gradient-type lensing signal. As pointed out in \cite{Pitrou:2015iya}, the five degrees of freedom, given by
two anisotropic expansion coefficients (e.g.\ $\beta_1$ and $\beta_2$) and the element of $SO(3)$ which specifies the coordinate frame $(\mathbf{\hat{x}}_1, \mathbf{\hat{x}}_2, \mathbf{\hat{x}}_3)$,
can be worked out from the five multipole coefficients $a_{2m}$ of the quadrupole.

The amplitude of this lensing due to an anisotropic spacetime could be used to derive constraints on the spacetime shear. However since the expansions coefficients $\beta_i(\chi)$ are  functions of time one has to specify a model for the anisotropies and derive the $\beta_i(\chi)$ consequently, as, e.g.~\cite{Adamek:2011pr} or anisotropic dark energy\footnote{One can modify the dark energy equation of state $p_{ij}= (\omega \, \delta_{ij}+\delta\omega_{ij})\rho_{\Lambda}$, to include an  anisotropic stress $\Pi_{ij}= \rho_\Lambda \, \delta\omega_{ij}$.}. In Ref.~\cite{Pereira:2015jya} (or Ref.~\cite{Pitrou:2015iya} for more details) the authors point out that a promising signal to constrain the anisotropies is not the lensing signal due to anisotropic expansion directly but  the fact the lensing signal form scalar perturbations evolving in an anisotropic universe has a large scales B-mode.  In these references the following constraints are derived:
$$  \frac{\si_0}{H_0} < 0.01\,,$$
where $\si= \sqrt{\si_{ij}\si^{ij}}=2\sqrt{\sum_i\dot\beta_i^2}$ is the shear of the constant time hypersurfaces and the index $0$ denotes present time.
This constraint will be achieved by future large scale structure surveys like Euclid or SKA (the Square Kilometer Array). The limit obtained from present data is only about $\si_0/H_0<0.4$.

\section{Conclusions}\label{s:con}
In this paper we have computed the E-type and B-type lensing potentials from tensor perturbations. We have applied our findings to determine the effect of lensing from tensor perturbations on the CMB. We have considered two applications, primordial  gravitational waves from inflation and the scalar-induced tensor perturbations. While the former are hypothetical with unknown amplitude (only upper limits exist), the latter can be computed from the measured scalar perturbation amplitude and spectrum  to second order in perturbation theory and fully with numerical simulations. We have also stressed that unlike inflationary gravitational waves, scalar-induced tensor fluctuations are not free waves. Most of their amplitude is not oscillating but represents a tensor anisotropy of spacetime. Finally we have used the formalism developed in this work to compute the linearized lensing potential of an anisotropic, Bianchi I spacetime. Comparing our derivation with the one given in Ref.~\cite{Pitrou:2015iya} gives an impressive demonstration  of the usefulness of our formalism. We have seen that the effect from tensor perturbations is most probably too small to be detected in the CMB, but it might be promising to apply the results found here to compute the effect of lensing by  tensor perturbations on cosmic shear~\cite{Schmidt:2012nw}, the tidal field~\cite{Schmidt:2013gwa} or on galaxy number counts~\cite{Montanari:2015rga}.

\FloatBarrier

\subsection*{Acknowledgments}
This work is supported by the Swiss National Science Foundation.

 \appendix
 
 \section{Euler Rotation} \label{appA}
 
    \begin{figure}[h]	
\centering
    	\includegraphics[scale=0.23]{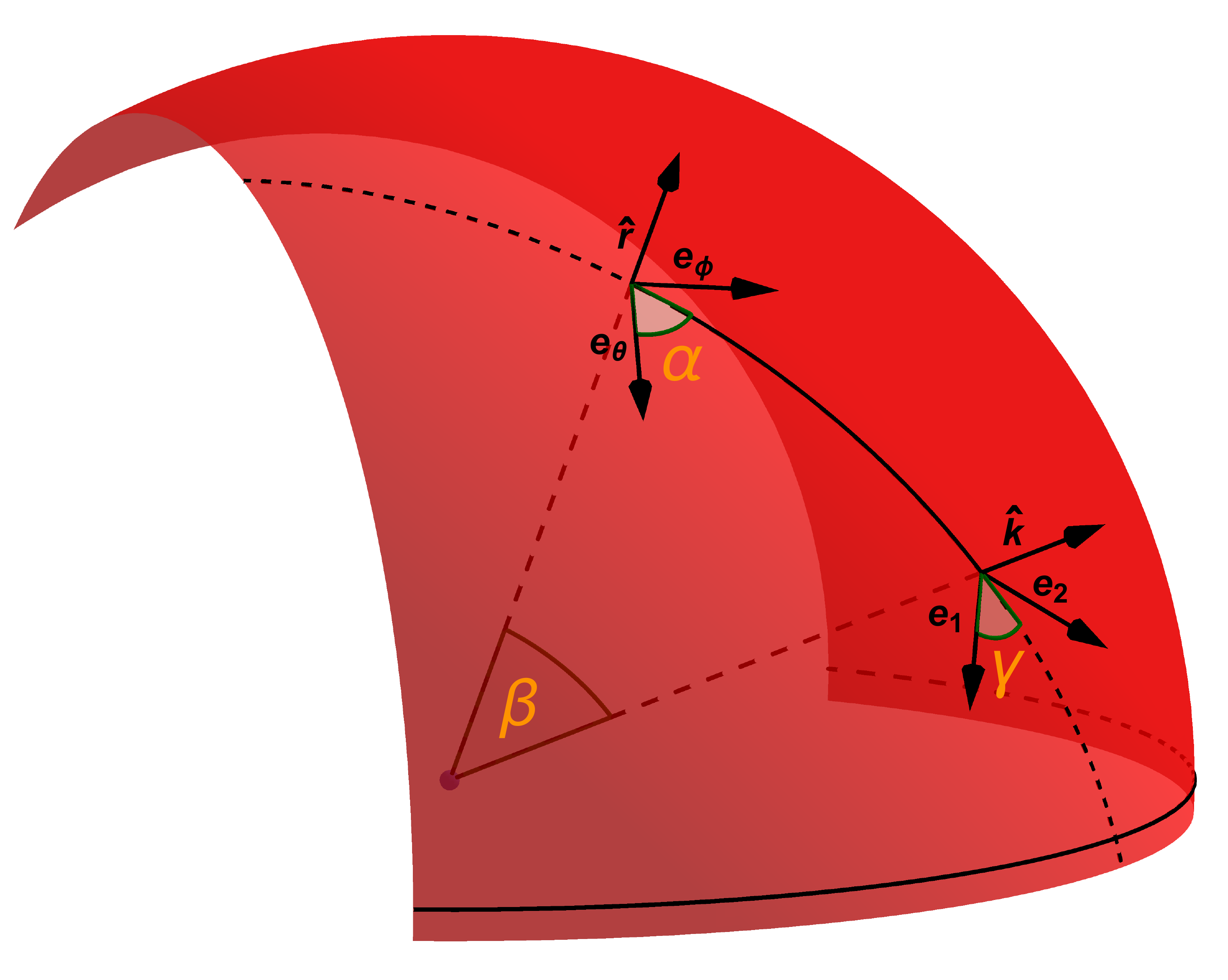}
	\caption{The two vectors $\hat{\mathbf{k}}$ and $\mathbf{n}$, together with the three Euler angles ($\alpha,\beta,\gamma$) necessary to rotate ($\mathbf{e}_1,\mathbf{e}_2,\hat{\mathbf{k}}$) into ($\mathbf{e}_{\theta},\mathbf{e}_{\phi},\mathbf{n}$).}
   	 \label{fig:euler}
\end{figure}
 
In Section~\ref{ss:1.1} we derived the displacement vector in eq.~(\ref{eq:defang}). We write it in Fourier components as
 
 \begin{equation}
\alpha_i= \int_{0}^{\chi_*} d\chi \, \int \frac{d^3 \mathbf{k}}{(2 \pi)^3} \, \left[ \frac{h^{ri}_{\mathbf{k}} (\chi,\theta_0,\phi_0)}{\chi} e^{-i \mathbf{k \cdot x}}+ \frac{1}{2} \frac{\chi-\chi_*}{\chi \chi_*} \nabla_i (h^{rr}_{\mathbf{k}}(\chi,\theta_0,\phi_0)e^{-i \mathbf{k \cdot x}}) \right] .
\label{eq:defangfour}
\end{equation}
 
 Since the power spectrum is usually written in terms of the two polarization modes $h_{\mathbf{k}}^{\oplus,\otimes}$ or in terms of the two helicity modes $h_{\mathbf{k}}^{\pm}$, we need to transform the Fourier components of the perturbation from the basis $(\mathbf{n},\mathbf{e}_{\theta},\mathbf{e}_{\phi})$, namely $\mathbf{h}^S_{\mathbf{k}}$, into the Fourier components of the basis $(\hat{\mathbf{k}}, \mathbf{e}_1, \mathbf{e}_2)$, namely $\mathbf{h}_{\mathbf{k}}$. In order to do that we perform a rotation with Euler angles ($\alpha,\beta,\gamma$) to rotate $(\theta_k,\phi_k)$ into our direction of observation $(\theta,\phi)$. This rotation is performed at fixed $\hat{\mathbf{k}}$, for every $\hat{\mathbf{k}}$, so that $\alpha$, $\beta$ and $\gamma$ are function of both $\hat{\mathbf{k}}$ and ${\mathbf{n}}$.
 The rotation proceeds as follows (see Figure~\ref{fig:euler}):
 \begin{enumerate}
 \item{Rotation around $\hat{\mathbf{k}}$ by angle $\gamma$ to align $\mathbf{e}_1$ with the \emph{node line} connecting $\hat{\mathbf{k}} $ and $\mathbf{n}$}
 \item Rotation around $\mathbf{e}_2$ by angle $\beta$ to align $\hat{\mathbf{k}}$ with $\mathbf{n}$
 \item Rotation around $\hat{\mathbf{k}}= \mathbf{n}$ by angle $\alpha$ to align $\mathbf{e}_1$ with $\mathbf{e}_{\theta}$ and $\mathbf{e}_2$ with $\mathbf{e}_{\phi}$.
 \end{enumerate}
The transformation is given by

 \begin{equation}
 \mathbf{h}^S_{\mathbf{k}} = R_1(-\alpha)R_3(-\beta)R_1(\gamma) \, \, \mathbf{h}_{\mathbf{k}} \,  \, R_1^T (\gamma) R_3^T(-\beta) R_1^T(-\alpha) .
 \end{equation}
  
 The result of this rotation is given in eq.~(\ref{eq:hsphetomod}). We can now write eq.~(\ref{eq:defangfour}) in terms of the helicity components $ \alpha_\pm \equiv \frac{1}{\sqrt2} (\alpha_\theta \mp i \alpha_\phi)$, as
  
  \begin{equation*}\begin{split}
  \alpha_\pm &= \frac{1}{2 \sqrt{2}} \int_{0}^{\chi_*} d\chi \int \frac{d^3 \mathbf{k}}{(2 \pi)^3} \Bigg[ \, \frac{1}{\chi} \, \Bigl( \sin \beta  \left((\cos \beta \mp 1) e^{2 i \gamma }  \; h_{\mathbf{k}}^+  + (\cos\beta  \pm 1) e^{-2 i \gamma } \; h_{\mathbf{k}}^-  \right) \Bigr) \, e^{i \alpha} \, e^{i k \chi \cos\beta} \\
  &+ \frac{\chi - \chi_*}{2 \, \chi \, \chi_*} \, (\nabla_\theta \mp i \nabla_\phi) \Bigl( \sin^2 \beta \, (  e^{2 i \gamma} \: h_{\mathbf{k}}^+ +  e^{-2 i \gamma} \: h_{\mathbf{k}}^-) \, e^{i k \chi \cos\beta} \Bigr) \Bigg] ,
  \end{split} 
  \end{equation*}
where we used the fact that $\beta$ is the angle between $\hat{\mathbf{k}}$ and $\mathbf{n}$. 
In this way the solutions of eq.~(\ref{eq:system}), namely the lensing potentials, are written in terms of the helicity modes and the Euler angles. Once the dependence on the Euler angles is eliminated with eq.~(\ref{eq:Eulspherical}) and the angular dependence on $(\theta,\phi)$ recast into the $Y_{\ell m}$s, the harmonic expansion coefficients $a_{\ell m}$ and $b_{\ell m}$ are only functions of $k, \chi$ and $h^\pm_{\mathbf{k}}$ and we finally obtain the expressiomns~(\ref{eq:scalarpotential}) and~(\ref{eq:Clomega}) for their spectrum.

  \section{Scalar-induced tensor perturbations} \label{appB}
  
  We briefly summarize the results of \cite{Ananda:2006af,Baumann:2007zm} on the tensor spectrum $P_h^{(2)}(k,t)$ induced by scalar perturbations at second order. We start with a metric perturbed at second order:
  
  \begin{equation*}
  \td s^2= a^2(t) \left[-(1+2\Phi +2\Phi^{(2)}) \td t^2 + 2V_i^{(2)} \td t \, \td x^i +\left( (1-2\Psi -2 \Psi^{(2)})\delta_{ij} +h_{ij} \right) \td x^i \td x^j    \right] \;,
  \end{equation*}
where $\Phi$, $\Psi$ are the Bardeen potentials and we ignore primordial gravitational waves so that $h_{ij}= h_{ij}^{(2)}$. We want to compute the tensor perturbations sourced at second order so that we need to consider:
 
 \begin{equation}
 G_{ij}^{(2)}= 8 \pi G \, T_{ij}^{(2)} ,
 \label{eq:einsten2}
 \end{equation}
where $G_{ij}^{(2)}$ is the second order spatial Einstein tensor and
  
  \begin{equation}
  {T^{(2) \,i}}_j= (\rho +P)v^{(1) \, i}{v^{(1)}}_j+ P \, {\Pi^{(2) \, i}}_j+ P^{(1)} {\Pi^{(1) \, i}}_j +P^{(2)} \delta^i_j
  \label{eq:t2}
  \end{equation}
is the second order energy momentum tensor with $\rho, P, v, \Pi$ the energy density, pressure, velocity and anisotropic stress, respectively. Acting on eq.~(\ref{eq:einsten2}) with $\Lambda_{ij}^{\ell m}$, the projection tensor constructed with the polarization tensors of eq.~(\ref{eq:polten}) that extracts the transverse, traceless part of any tensor, we get
  
  \begin{equation}
  \ddot h_{ij} + 2 \HH \dot h_{ij} - \nabla^2 h_{ij} = -2 {\Lambda_{ij}}^{\ell m} S_{\ell m} ,
  \label{eq:eom2}
  \end{equation}
where $S_{\ell m}$ is the source term as a function of $\Phi$, $\Psi$, $\omega=P/\rho$ and $c_s^2= P^{(1)}/\rho^{(1)}$. In Fourier space the equation of motion~(\ref{eq:eom2}) is written
  
\begin{equation}
\ddot{h}_k +2 \HH \, \dot{h}_k+k^2 h_k= S(k,t) .
\label{eq:eomk2}
\end{equation}
  
 The source term $S(k,t) $ is, as stated before, a convolution of two scalar perturbations at different wave numbers. A lengthy calculation yields
 
 \begin{equation*}
 \begin{aligned}
S(k,\chi) &= 8 \int \frac{\td^3 \bq}{(2 \pi)^3} \, \mathbf{e}^{\oplus \, \ell m}(\mathbf{k}) \, q_\ell q_m \Biggl[ \left(\frac{7+3 \omega}{3(1+\omega)}-\frac{2 c_s^2}{\omega} \right) \, \Phi_{\bq}(t) \Phi_{\bk-\bq}(t) + \left( 1-\frac{2 c_s^2 q^2}{3 \omega \HH^2} \right) \Psi_{\bq}(t) \Psi_{\bk-\bq}(t) \\
& +\frac{2c_s^2}{\omega} \left( 1+ \frac{q^2}{3 \HH^2} \right) \Phi_\bq(t) \Psi_{\bk-\bq}(t)+ \left(\frac{8}{3(1+\omega)}+\frac{2c_s^2}{\omega} \right) \frac{1}{\HH} \Phi_\bq(t) \dot \Psi_{\bk-\bq}(t)\\
& -\frac{2 c_s^2}{\omega \HH} \Psi_\bq(t) \dot \Psi_{\bk-\bq}(t) + \frac{4}{3(1+\omega)\HH^2} \dot \Psi_\bq(t) \dot \Psi_{\bk-\bq}(t) \Biggr] .
 \end{aligned}
 \end{equation*} 
  
 Solutions of eq.~(\ref{eq:eomk2}) can be found using the Green's function method: given $h^1_\bk (t)$ and $h^2_\bk (t)$ solutions of the homogeneous equation
  
  \begin{equation}
\ddot{h}_k +2 \HH \, \dot{h}_k+k^2 h_k= 0 .
  \label{eq:eomhom}
  \end{equation}
we can construct the Green's function for eq.~(\ref{eq:eomk2})
  
    \begin{equation}
    g_k(t,t')=\theta(t-t') \, \frac{h^1_k (t) h^2_k (t')-h^1_k (t')h^2_k (t)}{\dot h^1_k (t') h^2_k (t')-h^1_k (t') \dot h^2_k (t')} ,
  \label{eq:green}
  \end{equation}
which solves the equation $\ddot{g}_k +2 \HH \, \dot{g}_k+k^2 g_k= \delta(t-t')$. The solution of~(\ref{eq:eomk2})  is now given by
  
  \begin{equation}
  h_k(t)= \int \td t' g_k(t,t') \, S(k,t') ,
  \end{equation}
so that, for the power spectrum, we have

\begin{equation}
P_h^{(2)}(k,t) \propto \braket{h_\bk(t) h^*_\bq(t)}= \int \td t' \td t'' \; g_\bk(t,t') \, g_\bq^* (t,t'') \braket{S_\bk(t') S_\bq^*(t'')} .
\end{equation}
  
In the matter dominated era and in the radiation dominated era the homogeneous solutions of eq.~(\ref{eq:eomhom}) are given by combinations of $\sin(x)$ and $\cos(x)$ with $x= k t$. When dark energy is relevant the homogeneous solutions have to be determined numerically.  
  
 \newpage

\bibliographystyle{utphys}
\bibliography{GW-refs}

\end{document}